# Nuclear Magnetic Resonance Study of Monoclonal Antibodies Near an Oil-Water Interface

Jamini Bhagu,[1,2] Lissa C. Anderson[2,3], Samuel C. Grant,[1,2] and Hadi Mohammadigoushki[1, 2, a]

[1)]*Department of Chemical and Biomedical Engineering, FAMU-FSU College of Engineering, Tallahassee, FL, USA, 32310*

[2)]*Center for Interdisciplinary Magnetic Resonance, National High Magnetic Field Laboratory, Florida State University, Tallahassee, FL, USA, 32310*

[3)]*Department of Chemistry and Biochemistry, Florida State University, Tallahassee FL, USA, 32306*

Monoclonal antibodies (mAb) represent an important class of biologic therapeutics that can treat a variety of diseases including cancer, autoimmune disorders or respiratory conditions (e.g. COVID-19). However, throughout their development, mAb are exposed to air-water or oil-water interfaces that may trigger mAb partial unfolding that can lead to the formation of proteinaceous aggregates. Using a combination of dynamic surface tensiometry and spatially resolved 1D $^1$H NMR spectroscopy, this study investigates if adsorption of a model IgG2a-κ mAb to the oil-water interface affects its structure. Localized NMR spectroscopy was performed using voxels of 375 µm, incrementally approaching the oil-water interface. Dynamic interfacial tension progressively decreases at the oil-water interface over time, confirming mAb adsorption to the interface. Localized NMR spectroscopy results indicate that, while the number of mAb-related chemical resonances and chemical shift frequencies remain unaffected, spectral line broadening is observed as voxels incrementally approach the oil-water interface. Moreover, the spin-spin ($T_2$) relaxation of the mAb molecule was measured for a voxel centered at the interface and shown to be affected differentially across the mAb resonances, indicating a rotational restriction for mAb molecules due to presence of the interface. Finally, the apparent diffusion coefficient of the mAb for the voxel centered at the interface is lower than the bulk mAb. These results suggest that this specific mAb interacts with and may be in exchange with bulk mAb phase in the vicinity of the interface. As such, these localized NMR techniques offer the potential to probe and quantify alterations of mAb properties near interfacial layers.

---

[a]Corresponding Authors; hadi.moham@eng.famu.fsu.edu and grantsa@eng.famu.fsu.edu

## I. INTRODUCTION

Monoclonal antibodies (mAb) represent an important class of biologic therapeutics that are used in the treatment of a variety of diseases including cancer, asthma, infectious diseases, cardiovascular diseases and autoimmune disorders[1–5]. Despite their enormous lifesaving potential, a critical challenge is to keep mAb stable during molecular development and processing. During processing, transportation and administration, mAb frequently are exposed to hydrophobic interfaces in the form of oil-water in syringes (for lubrication purposes) and air-water (found in the headspace in vials or as entrained bubbles)[6–10]. mAb are amphiphilic molecules that readily adsorb to such hydrophobic interfaces. mAb adsorption at these hydrophobic interfaces may lead to partial unfolding and structural changes, and thus, their shelf-life, quality, safety and bioactivity may be compromised[8]. Therefore, probing the *in-situ* structure of mAb at hydrophobic interfaces is a crucial step to their optimal processing and a better understanding of how to mitigate these risks during molecule development. As a result, considerable interest has been devoted to investigate the adsorption of pure mAb to hydrophobic interfaces[11–15], effects of anionic surfactants on mAb adsorption to hydrophobic interfaces[15–17] and the impacts of surface flows on mAb adsorption to hydrophobic interfaces[17–21].

Recently, a number of techniques have been employed to study adsorption of pure mAb at hydrophobic interfaces. These approaches include confocal microscopy of labeled mAb[11], interfacial ellipsometry[14], x-ray or neutron reflectivity[13–15,22,23]. Using these techniques, researchers have measured the total interfacial area covered by mAb, and the thickness of the interfacial layer at hydrophobic interfaces. In particular, using neutron reflectivity (NR), Smith *et al.*[15] and Li *et al.*[13] studied adsorption of a monoclonal antibody (named as COE-3) at the air-water interface. These researchers showed that by increasing COE-3 concentration in the bulk, the amount of COE-3 adsorption at the interface increases, while the thickness of the adsorbed layer remains unchanged. It is expected that by increasing the mAb concentration in the bulk, the adsorption to the interface also increases until the interface is fully covered by the mAb. Because the thickness of the interfacial layer and size of the mAb molecules were the same (≈ 100 Å), these authors concluded that mAb retained their globular shape as they adsorbed to the interfaces. The latter observation is in contrast to that of Leiske *et al.*[11] who postulated that mAb not only unfold and potentially denature (*i.e.*, change their native structure) at the air-water interface, but also that the mAb layer at the air-water interface is much thicker (≈ 5 $\mu$m[11]). Leiske *et al.*[11] used Nile red to probe the hydrophobicity of mAb at the air-water interface via a fluorescence microscopy technique, and showed that the hydrophobicity of mAb increases upon adsorption at the air-water interface. The increase in mAb hydrophobicity was linked to unfolding and changes in their structure[11]. Additionally, Couston *et al.*[24] showed that depending on solution pH, and mAb concentrations, mAb may form multi-layers with a thickness as large as 400 Å at hydrophobic interfaces for mAb concentrations around 5 mg/mL. More recently, Kalonia *et* *al.*[25] investigated the adsorption of the NIST mAb at solid-liquid interfaces across a wide range of mAb concentrations (0.1–100 mg/mL). Using Quartz Crystal Microbalance, these researchers demonstrated that at higher concentrations, mAbs form multiple layers at the interface. The thickness of the layer closest to the interface was observed to vary between 5–8 nm as the mAb concentration increased. Interestingly, a much thicker interphase, approximately 270 nm in thickness, was measured for the second (outer) layer.

On the other hand, studies on mAb adsorption to the oil-water interface are sparse[26–29]. Randolph and co-workers studied adsorption of an IgG1 solution at a silicone oil-water interface via a size exclusion chromatography[26] and microscopy[27], and showed that mAb at the oil-water interface experience aggregate formation and possibly denaturation. In addition, Ruane *et al.*[28] investigated adsorption of COE-3 to the silicone oil-water interface via NR and measured the thickness of the adsorbed layer. Similar to their previous work for the air-water interface[15], Ruane *et al.* showed that the thickness of the adsorbed layer is similar to the size of the individual mAb molecules, and concluded that mAb retain their globular structure at the silicone oil-water interface[28]. Note the thickness of the adsorbed layer may not be a suitable criterion for inferring mAb structure at hydrophobic interfaces because mAb could deform at the interface but still retain the same thickness. Therefore, it is still an open question whether structure of the mAb molecule is modified upon adsorption to hydrophobic interfaces.

Determining the *in situ* structure of mAb at hydrophobic interfaces have remained a critical challenge due to limitations of the traditional techniques used to probe mAb adsorption to hydrophobic interfaces. One of the profoundly underused analytical techniques for studies of mAb structure and adsorption at hydrophobic interfaces is solution-state Nuclear Magnetic Resonance (NMR) spectroscopy. NMR exhibits remarkable sensitivity to even subtle alterations in protein structure because proton chemical shifts, lineshapes and relaxations are sensitive to the spatial arrangement of amino acids of proteins. A limited number of experiments have used NMR spectroscopy to study and compare different mAb formulations in the bulk solution[30–36]. As an example, Poppe *et al.*[30] used 1D $^1$H NMR spectroscopy as a tool to evaluate the similarities between two IgG1 antibodies using a protein fingerprint by a lineshape enhancement method. In this approach, the NMR spectra of various mAb in the bulk solution are compared to each other using a cross-correlation coefficient[30]. Other researchers have used $^1$H,$^{13}$C 2D gradient-selected heteronuclear single quantum coherence (gHSQC) NMR spectroscopy to compare different mAb formulations in the bulk of aqueous solutions[33].

Despite these advances, NMR-based methods have not been used to assess the impact of mAb adsorption to hydrophobic interfaces on mAb structure. In this study, a spatially resolved NMR localization method is employed for the first time to probe the impact of interfacial adsorption on mAb structure. In particular, spatially and frequency selective 1D $^1$H NMR spectra, chemical shifts, $T_2$ (spin-spin) relaxation and self-diffusion coefficient of mAb exposed to an oil-water interface were measured. The spatial resolution of a localized $^1$H NMR spectroscopy and diffusometry is limited by magnetic field gradient strength and the signal-to-noise ratio (SNR) of the NMR signal[37,38]. In this study the SNR is critical because the concentration of the mAb is in the semi-dilute regime, and generally SNR is low. As will be discussed in this paper, the smallest voxel size that generates a reasonable mAb SNR at the interface is 375 µm. Therefore, any acquired NMR data from the voxel centered at the oil-water interface will be associated with both interfacially adsorbed mAb layers and bulk mAb near the interfacial layer.

Note that at low mAb concentrations (e.g., up to 1 mg/mL), the interfacial thickness is expected to be much smaller than the voxel size. However, at higher concentrations (*e.g.*, > 5 mg/mL), the formation of an interphase and multilayers of adsorbed mAbs, with a total thickness in the submicron range, make the NMR based measurements relevant. As noted by Kalonia *et al.*[25] when very thick (>100 nm) multi-layers are present, it is challenging to capture them through neutron reflectivity because of weak contrast and high uncertainty at high wave-vector values. Furthermore, previous studies have shown that mAb adsorbed at the solid-liquid interface may experience a dynamic exchange with the bulk mAb, which was evidenced by formation of sub-visible (≈1-10 µm) aggregates of mAb in the bulk solution near the interface[25,39]. Therefore, acquiring NMR-based data from a voxel that size may shed light on the effect of adsorption on mAb structure, possible formation of multilayers, and exchange between the oil-water interface and bulk in the vicinity of the interfacial layer at concentrations that are relevant to pharmaceutical formulations, where it is often difficult to assess the formation of an interphase with other analytical techniques. To acquire a multi-scale understanding of the dynamics of mAb at this oil-water interface, NMR-based measurements were supplemented with dynamic surface tensiometry measurements to validate the interfacial activity of the mAb to the oil-water interface.

Pharmaceutical formulations have several excipients including surfactants, different buffers, etc[40,41]. Some of these excipients provide an additional level of complexity to the mAb system and their nature of interaction with mAbs may further complicate the picture, while also contributing additional NMR resonances. One example of these excipients is surfactants. Several works suggested that surfactants may bind to mAb and perturb (unfold) their native structures[42–44]. In this study, the aim is to assess the nature of interactions between pure mAb and the oil-water interface to better understand how mAb themselves behave at the oil-water interface. Therefore, the mAb solution is analyzed without surfactants or other excipients.

## II. EXPERIMENTAL METHODS AND APPROACHES

### A. Materials

The model mAb considered for this study is a murine anti-maltose binding protein (MBP)

IgG2a-κ mAb that was produced in-house using a hybridoma technique at the FSU Department of Biological Sciences. The resulting supernatant containing mAb was purified using Ag/Ab protein column purification (21030, ThermoFisher Scientific, Waltham MA). The mAb sample was prepared at a concentration of approximately 36.4 mg/mL in 20-mM sodium phosphate buffer solution in deionized water. The concentration of mAb in commercial formulations varies from 10-200 mg/mL[40,41]. Therefore, the mAb concentration range is well within that used in the prior studies. The pH of the buffer solution is 7.4, which is close to the isoelectric point (pI) of the mAb used in this study. The pI of the mAb was calculated to be pH 8 by the Swiss Bioinformatics Resource Portal, Expasy[45,46], using the mAb sequence that was determined by mass spectrometry (see the mAb sequence and related information in the supplementary materials). mAb are known to be most interfacially active around pI[14]. In this work, the impact of pH on mAb adsorption or stability has not been investigated, choosing instead to focus on a pH very close to the pI at which this mAb is expected to be most interfacially active and at which conformational alterations would be most likely. As for the oil phase, a fluorocarbon-based oil (Fluorinert FC-43, 3M Corp, Mapplewood MN), which is biologically inert and contains almost no protons that could interfere with the proton NMR signal of the mAb, was used. The mAb solution and oil were placed in a 5-mm cylindrical bulb with a 50-mm stem custom-made by Wilmad-LabGlass (Vineland NJ). All experiments reported in this paper are performed at T = 30°C.

## B. Dynamic Surface Tensiometry

Dynamic surface tension measurement was used to confirm and assess the kinetics of molecular adsorption to the hydrophobic interfaces. The dynamic surface tension of mAb at oil-water interfaces was measured using a custom-made pendant drop method. As shown in Fig. 1, the setup consists of an inverted needle that is connected to a syringe pump, high speed camera, light source and temperature-controlled water bath. These experiments were performed by injecting a known volume of mAb droplet in a bath of oil. Interfacial adsorption is manifested by a change in the overall drop shape, which then can be correlated to changes in the surface tension. More detailed information about this setup and analysis can be found in our recent publication[47].

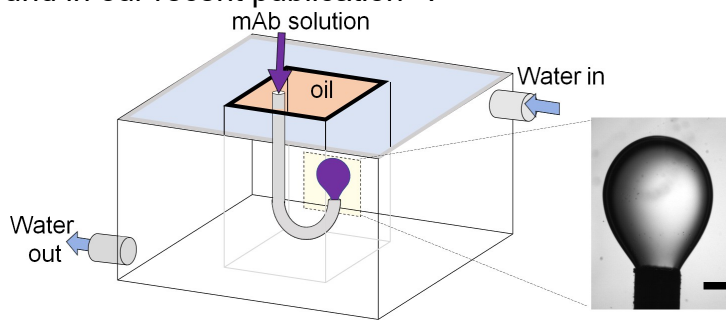

FIG. 1. A schematic of the pendant drop method used for measuring the dynamic surface tensions at the oil-water interface. Note that the oil phase is denser than the mAb solution, and therefore,

the needle is inverted. The oil chamber is placed inside a temperature-controlled water bath. The size of the scale bar in the image to the right is 1 mm.

## C. 1D $^1$H NMR Spectroscopy

All NMR scans were conducted using the 21.1-T (900-MHZ) vertical magnet at the National High Magnetic Field Laboratory. This magnet is equipped with Bruker Biospin (Billerica, MA) Neo Console and controlled using Paravision 360. A 10-mm linear birdcage RF coil (Bruker) and specially built gradient coil with a maximum gradient strength of $G_{max}$ = 60 G/cm/axis (RRI Inc., Billerica, MA) was used for these experiments. Although the RF coil and gradient system are adapted to the 21.1-T system and operation at the $^1$H Larmor frequency of 900 MHz, these components are commercially available for operation at lower magnetic fields equipped for microimaging applications.

Figure 2(a) presents a schematic of the measuring cell, illustrating how solutions containing mAb molecules and oil phases are arranged for localized NMR spectroscopy measurements. 1D $^1$H NMR spectrum was acquired in the bulk and at the hydrophobic interface using a spatially and spectrally selective relaxation-enhanced point resolved spectroscopy (REPRESS) sequence (shown schematically in Figure 2(b)). The REPRESS experiments were performed at a set of nine locations (shown as rectangles in Fig. 2(a)) axially shifted by 375 µm. Relaxation enhancement allows for frequency-selective suppression of water signal and results in higher sensitivity mAb signal through reduced magnetization transfer between the mAb and bulk water protons. A VAPOR water suppression was added before the REPRESS acquisition to eliminate any residual water signal[48]. In addition, a diffusion-weighted REPRESS sequence (DW-REPRESS, shown schematically in Figure 2(c)) was implemented using bipolar gradients applied on the y- and z-axes to measure diffusion within the bulk mAb and in the vicinity of the oil-water interface. The voxel sizes used in these experiments was 1.5×1.5×0.75 mm. These experiments were performed at two axial locations, one located in the bulk solution containing mAb far away from the interface, and one at the interface. The number of scans for each DW-REPRESS voxel was 2048 for mAb acquisitions but only 128 averages for water diffusion measurements. The acquired 1D NMR spectra were analyzed to assess changes in the mAb NMR chemical shifts, diffusivity and the $T_2$ (spin-spin) relaxation upon approach to the interface discussed in details below.

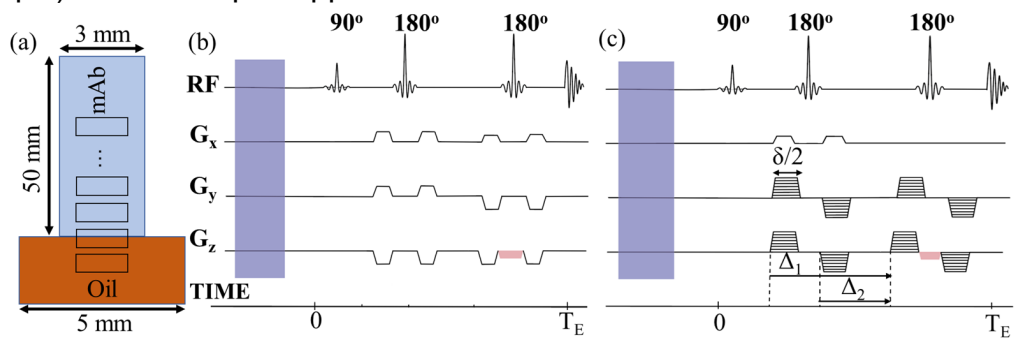

FIG. 2. (a) A schematic of the experimental cell used to investigate the effects of adsorption to oil-water interface on mAb structure. Pulse programs of the REPRESS (b) and DW-REPRESS sequences (c) used to obtain localized NMR spectra of the mAb solution. Blue regions in the sequences display the water suppression that was implemented before data acquisition. The gradient pulse (red) during the $2^{nd}$ $180°$ refocusing pulse is used for slice selection. Incremented gradients in (c) indicate the placement and timing of bipolar diffusion-weighting gradients, with $δ/2$ representing the gradient duration, $\Delta_1$ representing the duration between positive diffusion lobes, and $\Delta_2$ representing the duration between negative and positive diffusion lobes.

### D. NMR Data Analysis

The resulting NMR spectra of the mAb at different axial locations approaching the oil-water interface can be compared by cross-correlating their spectra. Poppe *et al.*[35] introduced a self-similarity index as SSI = 10× *log*(R/1-R) to evaluate the differences in the higher order
structures of various mAb formulations in bulk solutions. Here, *R* is the maximum value of the normalized cross-correlation function for the two NMR spectra. *R* varies as $0 \leq R \leq 1$, and with higher *R* and SSI, the more similar the structures of the mAb will be. Therefore, this technique was used to compare the 1D $^1$H NMR spectra of the bulk mAb to those voxels incrementally approaching the oil-water interface.

Additionally, by using the DW-REPRESS sequence, the self-diffusivity of the mAb molecules was measured in the bulk and in the vicinity of the oil-water interface. The normalized NMR signal intensity is directly linked to the molecular diffusion via[49,50]:

$$\frac{S(T_E)}{S(0)} = \exp(-BD) \qquad (1)$$

where $D$, $S(B)$ and $S(0)$ are apparent diffusion coefficient (or ADC), the intensity of the acquired NMR signal at different diffusion weightings ($B$) and NMR signal without weighting ($B=0$), respectively. $B = \gamma^2 G^2[\delta^2(\Delta_1 - \delta/3) - 2\delta(\delta/2)(\Delta_1 - \Delta_2 - \delta_1/2) + 2(\Delta_1 - \Delta_2 - \delta_1/2)(\delta/2)^2]$, where $\delta$ is the diffusion gradient duration ($\delta/2$ for each positive and negative lobe), $\gamma$ is the gyromagnetic ratio of the nuclei (e.g., for proton, $\gamma = 2.67 \times 10^8$ rad/s/T), $G$ is the magnitude of the magnetic gradient pulse, and $\Delta_1$ and $\Delta_2$ are the diffusion gradient separation times. For the experiments reported in this paper, $\delta = 6$ ms, $\Delta_1 = 12.5$ ms and $\Delta_2 = 7$ ms. The B values ranged from 92 – 10,000 s/mm$^2$. Because the size of the mAb molecules is much larger than water or other excipients present in the solution (e.g., salt or water), mAb ADC will be much smaller. Therefore, through diffusion-based filtering, chemical shifts associated with mAb molecules can be identified, and by fitting acquired NMR signals of the mAb to Eq. (1), the ADC for mAb molecules is determined.

Two relaxation mechanisms are involved in NMR experiments; $T_1$ (spin-lattice) and $T_2$ (spin-spin). While $T_1$ is the coefficient for regrowth of longitudinal magnetization to a thermal equilibrium value[51], $T_2$ relaxation is the process by which the transverse component of the magnetization decays[51]. Between these two relaxation mechanisms, $T_2$ is significantly sensitive to changes in viscosity, porosity and the molecular tumbling rate in proteins[52–54]. If any segments of the mAb molecule changes its tumbling rate upon adsorption to the oil-water interface, it will be detected in variations of the $T_2$ relaxation of the mAb. Therefore, $T_2$ relaxation data was acquired in the bulk, and results were compared to those measured for the voxel centered at the oil-water interface using the REPRESS sequence sampled at different echo times. The $T_2$ relaxation coefficient was obtained using the following equation:

$$\frac{S(T_E)}{S(0)} = \exp(-T_E/T_2), \qquad (2)$$

where $T_E$ is the echo time and $S(T_E)$ is the NMR signal intensity at a given echo time.

## III. RESULTS AND DISCUSSION

Prior to reporting the interfacial experiments with the mAb, we confirmed the nature of the mAb by performing SDS measurements (see Fig. S1 of the supplementary materials.) Based on these SDS-PAGE results, the molecular weights of the heavy and light chains are approximately 50 and 25 kDa, respectively. Consequently, these measurements imply an approximate molecular weight of 150 kDa for this specific mAb.

### A. Dynamic Surface Tension

Following the purification and characterization of the mAb, dynamic interfacial tension measurements were conducted using the pendant drop method. Figure 3 (a,b) presents a sequence of snapshots captured during the pendant drop experiments, illustrating the buffer-oil interface in (a) and mAb solution-oil interface in (b). In Figure 3 (c), the measured dynamic surface tensions for the buffer-oil and mAb-oil interfaces are depicted. At the buffer-oil interface, the interfacial tension remains constant, whereas at the mAb-oil interface, there is a gradual decrease over 300 s before reaching a stable level. The temporal variations in the dynamic surface tension of the mAb-oil interface are reflected in the evolving shape of the mAb droplet, as illustrated in Figure 3(b). These findings provide confirmation that mAb molecules are interfacially active and undergo adsorption at the oil-water interface.

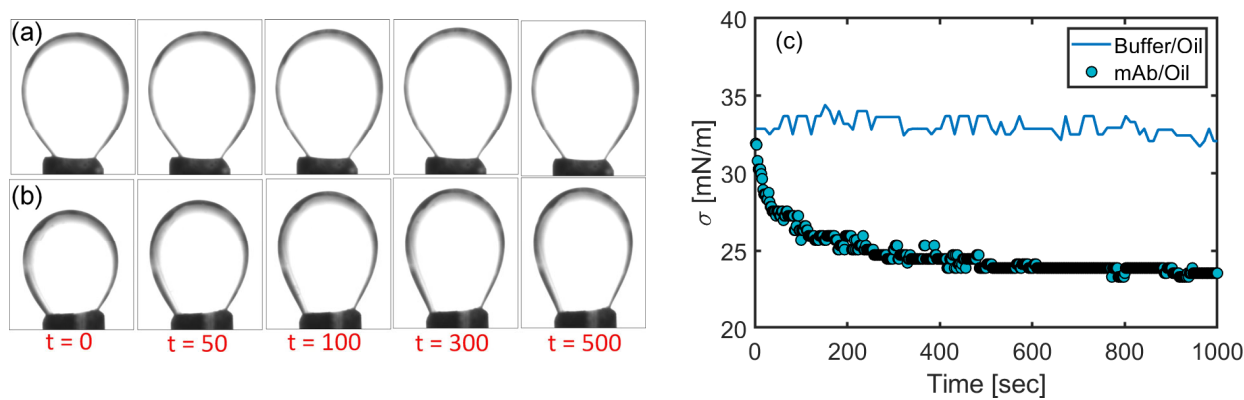

FIG. 3. A series of image sequences captured from pendant drop experiments illustrating the temporal evolution of the droplet of phosphate buffer (a) and mAb solutions (b) immersed in the oil phase. (c) The temporal evolution of the interfacial tensions at the oil-water interface for buffer- oil and mAb-oil interfaces.

## B.  1D $^1$H NMR Spectroscopy

Upon confirming adsorption of mAb to the oil-water interface, a volume-selective PRESS sequence was executed. Subsequently, $^1$H NMR spectra of the mAb were acquired in the bulk buffer solution, positioned away from the interface and at the oil-water interface. In Figure 4(a), the NMR spectra in the bulk mAb solution is depicted at various diffusion weightings, *B*. Interestingly, the NMR signal is dominated by the chemical resonance of the water molecules (here denoted at $\sigma$ = 4.7 ppm). Figure 4(b) shows the $^1$H NMR spectra for the voxel that is centered at the oil-water interface at various *B* values. The NMR spectra of water for the voxel centered at the interface exhibit a smaller signal-to-noise ratio and broader spectral lines compared to the bulk water data. Evidently, the dominance of the water peak (both in the bulk solution and in the voxel centered at the oil-water interface) is such that the remaining signal associated with the mAb is too weak to be resolved in these measurements.

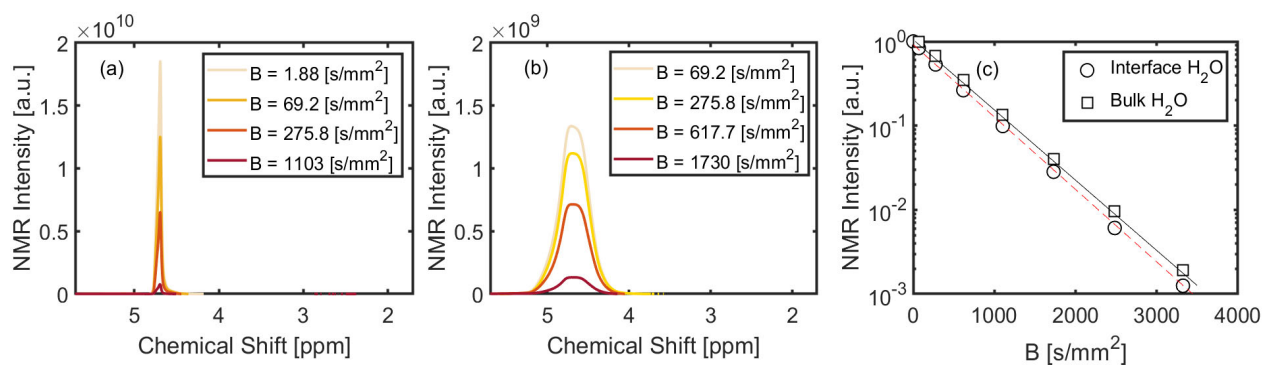

FIG. 4. 1D $^1$H NMR spectra of the mAb in the bulk solution (a) and in the voxel centered at the oil-water interface (b) obtained by a DW-REPRESS sequence without VAPOR water suppression. The signal attenuation for the water peak as a function of diffusion weighting both in the bulk and in the voxel centered at the interface (c). The dashed and dotted lines are the best fits to the experimental data. The diffusion coefficients for bulk water and interfacial water are $D = 1.98 \times 10^{-3}$ and $D = 1.93 \times 10^{-3}$ [mm$^2$/s], respectively.

The experimental data of water signal attenuation, in relation to diffusion weighting, are fitted to Eq. (1) to evaluate the apparent diffusion coefficient of the water molecules. In Figure 4(c), the signal attenuation is depicted as a function of diffusion weighting for water molecules in both the bulk and in the voxel centered at the oil-water interface. First, the resulting water diffusivity appears to be similar and does not seem to be influenced significantly by the presence of the interface. This result is not surprising as water molecules are not expected to behave differently near the oil-water interface. In addition, the diffusivity of the water in the mAb solution is slower than the water diffusivity reported for the pure water at 30°C (e.g., $D \approx 2.59 \times 10^{-3}$ [mm$^2$/s][55]). The reduction in water proton self-diffusivity is presumably caused by two factors. First, presence of the mAb could obstruct the self-diffusivity of the water molecules. Secondly, it is possible that a fraction of protons in the water is bound to mAb due to hydration. A similar hindered proton diffusion has been reported for water molecules in self-assembled surfactant solutions[49].

As mentioned earlier, the dominance of the unsuppressed water peak obscures any signals associated with the mAb molecules in both the bulk and in the voxel at the oil-water interface due to limitations in dynamic range. To address this issue, the REPRESS sequence employs a VAPOR pre-saturation module to suppress the water signal, allowing for the acquisition of locally resolved NMR spectra specifically associated with mAb molecules. In Figure 5(a), the $^1$H NMR spectra of the mAb solution at various distances away from the oil-water interface, acquired with the REPRESS sequence, are illustrated. Remarkably, the VAPOR REPRESS sequence completely suppresses the water signal, allowing for the resolution of other chemical resonances. In the bulk solution, three distinct chemical resonances are evident at approximately 3.9, 3.5 and 2.9 ppm. As proximity to the oil-water interface increases, these resonances do not shift but rather weaken and broaden. mAb typically contain both polar and non-polar amino acids including glycine, proline, lysine, arginine, glutamate and/or methionine. The three dominant resonances are near the reported chemical shifts for glycine, proline, lysine, arginine, glutamate and/or methionine side chains, and could be associated with each of these amino acids. For example, glycine expresses a chemical resonance around 3.56 ppm, the $H_\alpha$ protons in lysine and arginine side chains are reported to have chemical shifts in the range of 3.8-4.7 ppm[56] Additionally, the $H_{\varepsilon 2/\varepsilon 3}$ protons of the isopropyl side chain are reported to have chemical shifts between 2.7 and 3.1 ppm. Finally, the resonance at approximately 2.8 ppm could be associated with the $H_{\delta 2/\delta 3}$ protons of the arginine side chain, as their chemical shifts fall within the 2.8-3.4 ppm range[56]. In the oil phase, the signal eventually decays to zero. The latter result is expected because the selected oil neither has protons nor has penetration of the mAb beyond the interface, and therefore, no $^1H$ NMR signal is detected.

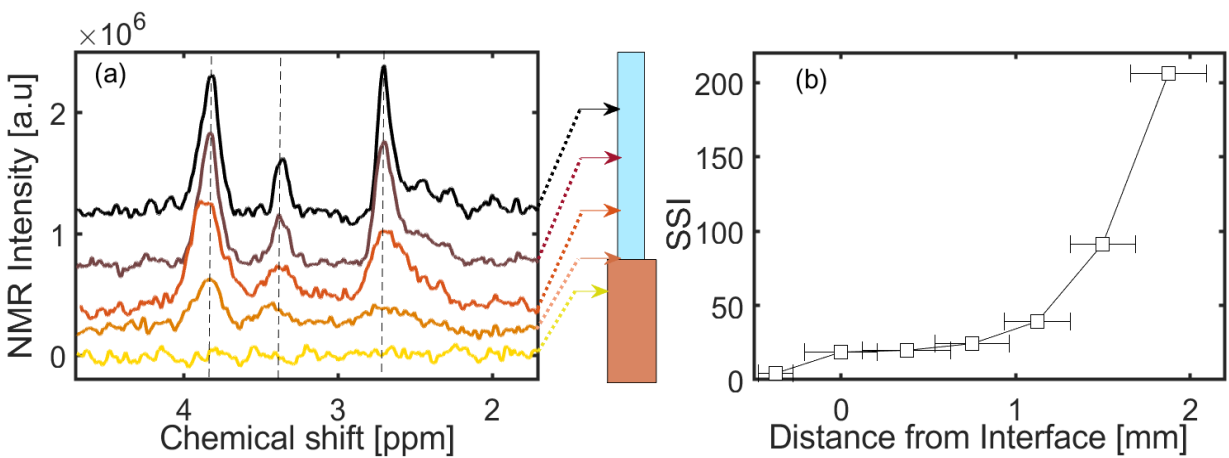

FIG. 5. $^1H$ NMR spectra of the mAb solution at different axial locations of the measuring cell. (a) From top to the bottom, the NMR spectra are acquired [1.875, 1.125, 0.375, 0 and -0.375] mm from the oil-water interface. NMR spectra are shifted vertically for visual clarity. (b) SSI for mAb spectra as a function of distance from the oil-water interface.

The initial step in evaluating the alteration in mAb structure upon interfacial adsorption involves employing the self-similarity analysis introduced by Poppe et al.[30]. This analysis considers the NMR spectra as analog signals that could be compared through a cross-correlation process. If NMR spectra change significantly, the changes should manifest in this scalar property. Figure 5(b) shows the resulting SSI obtained for NMR spectra of mAb shown in Figure 5(a). As the proximity to the oil-water interface increases, spectra exhibit a decreasing degree of self-similarity. Poppe et al. proposed that alterations in the SSI among different mAb formulations are indicative of variations in the mAb structure. In this context, the decline in SSI, linked to NMR signal broadening, may be attributed to two potential factors: alterations in the mAb structure with approach to the oil-water interface or magnetic susceptibility ($\chi$) effects. The latter effect arises from the mismatch in magnetic susceptibility between the bulk mAb solution and oil phase. This difference may distort the uniform magnetic field around the sample at the interface, consequently causing some signal broadening[57], though it should be noted that the oil used in this model system was chosen due to its lack of proton signal and a $\chi$ within 9% of water[58,59]. To distinguish between these two effects, a similar self-similarity analysis also was conducted for NMR spectra of water molecules (*i.e.*, Figure (4)). The bulk NMR spectra of water were compared to those of water located in the voxel centered at the oil-water interface (see Fig. S1 in the supplementary materials). Interestingly, the SSI for the water signal exhibits a comparable decline to that observed in Figure 5(b) for mAb molecules. Hence, even small magnetic susceptibility mismatches between the oil and water phases cannot be ignored with respect to the interpretation of the SSI analysis. To probe mAb interactions and alterations in the interfacial vicinity, additional quantitative MR metrics (i.e., chemical shift, diffusion and relaxation) were evaluated.

After conducting the self-similarity analysis presented in Figure (5), spatially-resolved, diffusion-weighted 1D $^1$H NMR spectroscopy was carried out on the mAb solution, both in the bulk and in the vicinity of the oil-water interface. Figure 6(a,b) show a series of NMR signals at the bulk (a) and in the voxel centered at the oil-water interface (b) for different diffusion weightings. In alignment with the localized NMR data presented in Figure 5(a), the chemical shifts associated with the mAb experience line broadening near the oil-water interface. In Figure 6(c), the normalized NMR signal attenuation is shown as a function of *B* values both in the voxel centered at the oil-water interface and in the bulk mAb solution. Note that the diffusivity obtained from each individual chemical resonance are the same, and the results reported in Figure 6(c) are the averaged signal attenuation for multiple replicates and three chemical resonances. The signal attenuation at both the bulk and in the vicinity of the interface are best fitted to a mono-exponential decay function (*i.e.*, Eq. (1)), highlighting the unrestricted (or free) self-diffusion of mAb molecules in this system. The diffusion coefficient of other commercially available mAb in the bulk solution has been reported before and is strongly dependent on the mAb type, the mAb concentration, the pH and the ionic strength of the solution[60,61]. The ADC of mAb measured in bulk solution at concentrations similar to the current study have been reported to be around $\approx 6 - 7 \times 10^{-11}$ m$^2$/s[32], and the current bulk mAb data is consistent with such diffusion coefficients.

Interestingly, the diffusion of mAb near the interface is significantly slower, nearly by a factor of 2, compared to mAb molecules in the bulk solution (see the calculated diffusion coefficients in the caption of Figure (6)). A slower diffusion of mAb measured in the voxel located at the oil-water interface suggests that mAb are strongly associated with each other at the oil-water interface, may form multi-layers, and possibly exchange with the mAb molecules that are in the vicinity of the interfacial layer. This finding is in line with previous studies that have shown that adsorption of mAb to solid-liquid interfaces is a dynamic process and involves exchange of adsorbed mAb with the bulk mAb as well as formation of thick mAb multilayers[13–15]. The exchange of the mAb between the bulk and solid-liquid interface was evidenced by formation of sub-visible particles in prior studies[25]. The spatial resolution of MRI is on the order 10μm, which is less than the typical range of aggregate sizes reported in the literature[25]. Therefore, the direct formation of such particles *in situ* cannot be assessed via MRI but alterations in mAb interaction in the vicinity of the interface can be inferred. It should be noted that previous studies flushed their cell because the interface was a solid-liquid interface, an approach that cannot be applied in the current oil-water phase system and would necessitate a major change in the design of the measuring cell. The reduction in the mAb diffusion coefficient could result from restrictions at the interface, interfacial exchange dynamics and/or formation of mAb aggregates. However, exchange or restriction are expected to yield a non-Gaussian diffusion profile[62]. With the mono-exponential or Gaussian diffusion decay evident for both interfacial water and mAb, it seems unlikely that either restriction or exchange would be major contributors to the slower diffusivity. As noted, the size of voxel in this study is 187 μm. The observed differences in mAb diffusivity near the interface suggest that mAb in this region may interact with each other over long distances, indicating that the influence of the interface extends far beyond a monolayer with a thickness of about 10 nm. This hypothesis is consistent with previous measurements on interfacial layer thickness measurements that at relatively high mAb concentrations (>1 mg/mL), mAb form multi-layers in the vicinity of the interfaces, where in addition to the interfacial layer, a much thicker interphase with thickness of approximately 200-300 nm is formed[25]. Therefore, it is surmised that the slower mAb diffusion near the interface is presumably due to the prevalence of mAb aggregates that can be detected in a voxel several times larger than the interfacial layer.

An additional important aspect of these measurements involves assessing the impact of interfacial adsorption on the $T_2$ relaxation of mAb molecules. Figure 7(a,b) show the NMR spectra of the bulk mAb and mAb in the vicinity of the interface at various echo times. Consistent with NMR spectra observed in previous measurements, three distinct chemical resonances are observed in Figure 7(a,b). It appears that the chemical frequencies remain similar and do not undergo a shift for mAb that are located in the voxel centered at the oil-water interface. Interestingly, among these three peaks, the $T_2$ relaxation appears to be significantly different. For the chemical shift closest to the water resonance, the signal decays water interface (b) at different diffusion weightings. The normalized signal attenuation for the chemical resonances as a function of B value (c). The continuous and dashed lines are the best fit of Eq. 1 to the experimental data. The apparent diffusion coefficients of mAb in bulk and in the vicinity of the interface are $D = 6.8 \times 10^{-5}$ and $D = 3.7 \times 10^{-5}$ [mm$^2$/s], respectively.

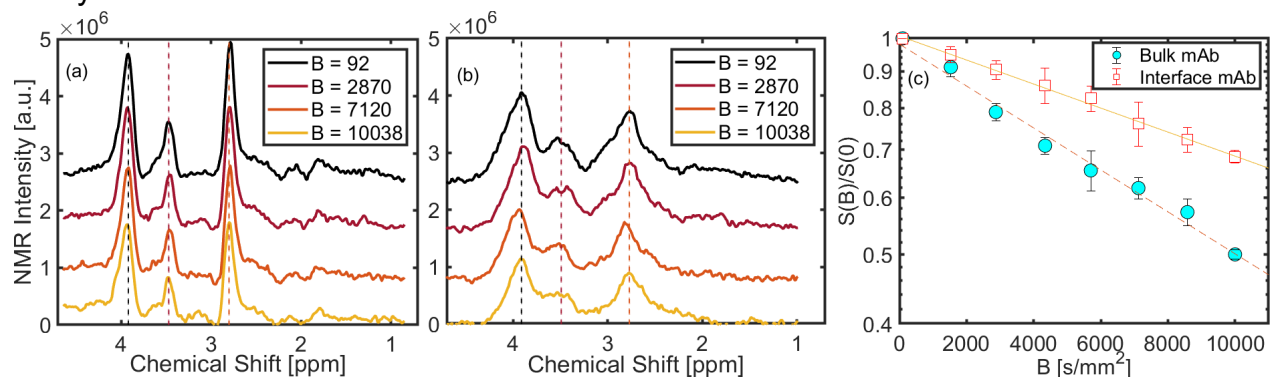

FIG. 6. 1D $^1$H NMR spectra of mAb solutions in the bulk (a) and in the vicinity of the oil-
13

much faster than the other chemical shifts, resulting in a lower $T_2$ value. For example, as $T_E$ changes from 16 to 65 ms, while the signal associated with $\sigma = 3.9$ ppm decays to zero, the other two chemical resonances $\sigma = 3.5$ and 2.9 ppm decay to much smaller extents.

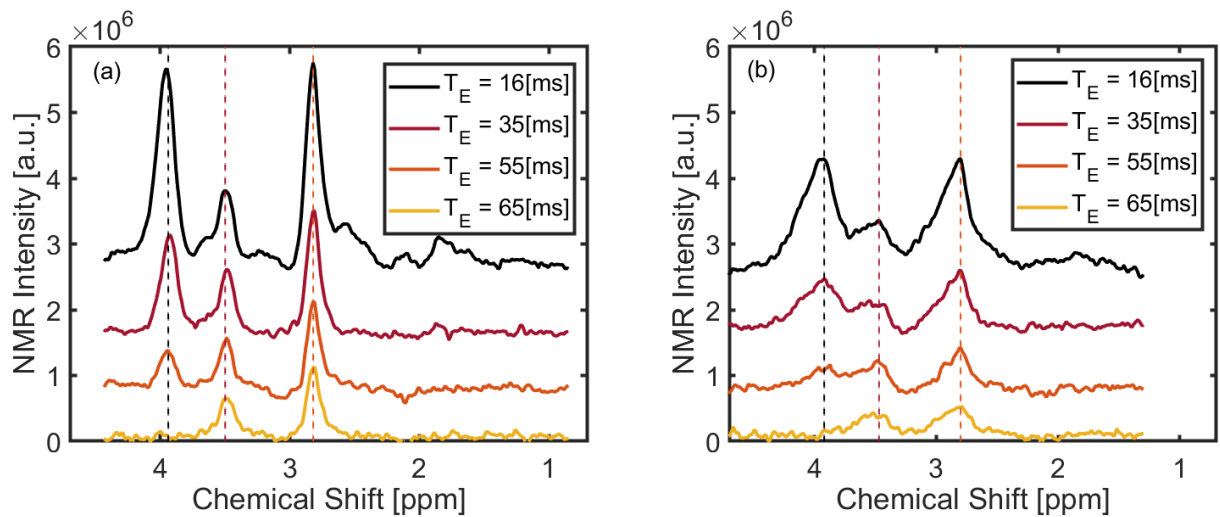

FIG. 7. 1D $^1$H NMR spectra for mAb in the bulk (a) and near the oil-water interface (b) at various echo times obtained via a REPRESS sequence.

Figures 8(a,b) depict the signal attenuation related to $T_2$ relaxation experiments for various chemical resonances, along with the best fit to Eq. (2). Figure 8(c) displays the corresponding $T_2$ coefficients for mAb in the bulk and mAb that are near the oil-water interface. The $T_2$ coefficients within the mAb molecule exhibit significant variations, indicating distinct mobilities across different segments of the molecule. Furthermore, the $T_2$ relaxation of each chemical resonance seems to be differentially affected by adsorption of the mAb to the oil-water interface. The lower $T_2$ values of the mAb in the vicinity of the interface for resonances at 2.9 and 3.5 ppm indicate that part of the mAb molecule is interacting with or rotationally restricted in the voxel centered at the oil-water interface, in partial agreement with slower diffusivities.

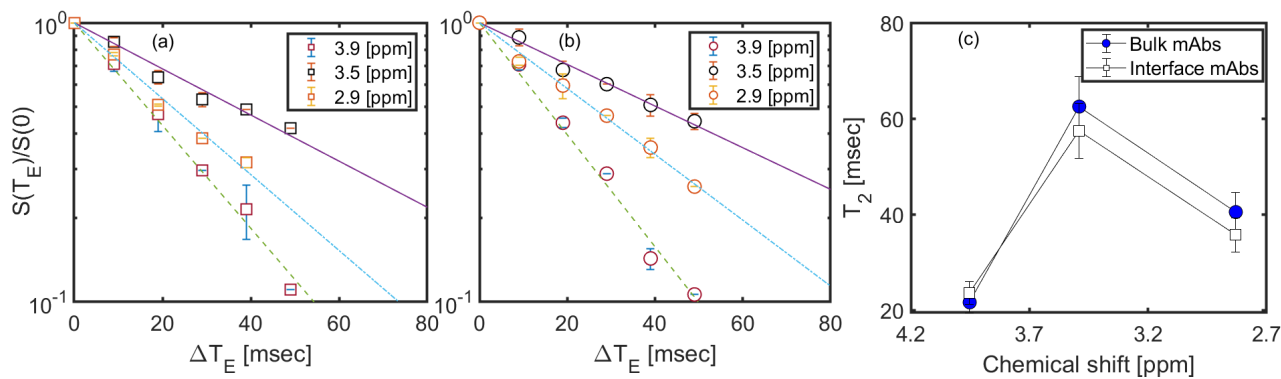

FIG. 8. The signal attenuation associated with different echo times for various chemical resonances in the bulk (a) and near the mAb/oil interface (b). The dashed lines are the best fits of Eq. 2 to the experimental data. Here $\Delta T_E = T_E\text{-}16$ [ms]. (c) $T_2$ or spin-spin relaxation of different chemical resonances of the mAb molecule. The circle and squares correspond to the bulk mAb and mAb in the vicinity of the interface, respectively.

Taken together, the findings of this study suggest that the number of mAb-related chemical resonances and chemical shift frequencies remain largely unaffected for this model mAb molecule with approach to the oil-water interface. However, findings from diffusion analysis and differential $T_2$ relaxation values for mAb resonances reveal that mAb exhibit a slower diffusion rate and interactions in proximity to the interface compared to their behavior in the bulk. These changes are consistent with increased association of the mAb molecules with each other and possible formation of small aggregates at the oil-water interface. The mAb aggregate then might undergo dynamic exchange with bulk mAb in the immediate vicinity of the interface, and form multilayers in the thick interphase between the bulk and oil-water interface. This thicker interphase may explain the slower diffusivity and a reduced $T_2$ relaxations over voxel volumes larger than the scale of the monolayer interfacial mAb layer. The latter statement is in line with current and prior measurements on different mAb at relatively high concentrations, suggesting that upon adsorption to hydrophobic interfaces, mAb tend to form multilayers with elevated viscosity[21,63,64] and that adsorbed mAb molecules to solid-liquid interfaces may exchange with the bulk solution[25,39]. To the best of our knowledge, a quantitative measurement of the rate and/or extent of exchange of the mAb molecules between oil-water interface and bulk mAb does not exist and may require isotope-labeling procedure similar to those reported for solid-liquid interfaces[65,66]. There are serious limitations that prohibit such measurements, including the complicated radioactive labeling technique and presence of a deformable oil interface that inhibit quantification of the extent and rate of exchange.

## IV. CONCLUSION

The first experimental investigation employing localized 1D $^1$H NMR spectroscopy to study the impact of interfacial adsorption of a model IgG2a-κ monoclonal antibodies in the vicinity of the oil-water interface was presented. The key outcomes of this investigation are outlined below.

First, dynamic surface tension measurements at the oil-water interface revealed a gradual decline toward a quasi-steady limit, providing confirmation of interfacial mAb adsorption to the oil-water interface. Secondly, the localized NMR spectra of the mAb molecules were tracked systematically in relation to their proximity to the oil-water interface. Although the chemical shifts remained consistent, the NMR spectra exhibited increased line broadening. Consequently, NMR spectra of the mAb became less self-similar as they approached the interface. This analysis suggests that the decline in the self-similarity index is correlated mostly with magnetic susceptibility for this model mAb at the oil-water interface. Thirdly, the $T_2$ relaxation coefficients of chemical shifts linked to mAb differ significantly, indicating distinct mobility in various parts of this mAb. However, it is noteworthy that the $T_2$ relaxation of each chemical shift appears differentially affected in the voxel centered at the oil-water interface, justifying a moderate alteration of rotational mobility near the interface. Finally, diffusion measurements demonstrated that mAb molecules exhibit a slower diffusion in areas near the oil-water interface. The latter results, taken with potential $T_2$ alterations, suggest the formation of associations (likely in the form of small aggregates of mAb that reside in a thick interphase area) at the oil-water interface that could be in significant exchange with the bulk mAb proximal to the interface.

As a final note, measurements within a voxel less than 200 $\mu$m (focusing only on the part of the voxel within the water phase) were able to probe mAb interactions with respect to MR metrics, demonstrating sensitivity to mAb-related alterations proximal to the interface. Future investigations will delve deeper into increasing the spatial resolution of the technique to understand how various parameters, such as the air-water interface, mAb type and concentration, and presence of surfactants or surface flows influence both 1D and 2D NMR fingerprints of mAb near hydrophobic interfaces.

## V. SUPPORTING INFORMATION

The mAb sequences, along with SDS-PAGE results, confirm the nature of mAb and a change in the self-similarity index (SSI) of water upon approach to the oil-water interface.

## VI. ACKNOWLEDGEMENTS

This work is partially supported through a grant from National Institute of Health (NIH) R21-AI163988. Part of this work was performed at the US National High Magnetic Field Laboratory (NHMFL), which is supported by the State of Florida and the National Science Foundation Divisions of Chemistry and Materials Research through Cooperative Agreement No.

DMR-1644779. In addition, HM gratefully acknowledges support by NSF through award CAREER CBET 1942150. SCG receives support from the NIH NINDS through R01-NS102395 and R01-NS072497.